\documentclass[preprint,10pt]{article}

\usepackage{amssymb}
\usepackage{amsmath}
\usepackage{graphicx,epsfig}
\usepackage{color}

\usepackage{lineno}

\begin{document}

%\begin{frontmatter}



%\title{Characterizing probability density distributions for
%domestic electricity profiles by using high-resolution electricity use data}

%\title{A probability distribution model of photovoltaic power
%production, household electricity use and electric vehicle home-charging}
%\title{On the probability distributions of photovoltaic power
%production, household electricity use and electric vehicle home-charging}
\title{On Non-Equilibrium Thermodynamics of Space-Time and
Quantum Gravity}


\author{Joakim Munkhammar \footnote{e-mail: joakim.munkhammar@gmail.com}}
%\author{Lars Mattsson$_b$}
%\author{Jesper Ryd\'{e}n$_c$}

%\address{a (Corresponding author) Built Environment Energy Systems
%Group, Department of Engineering Sciences, Uppsala University, SE-
%751 21 Uppsala, Sweden. e-mail: joakim.munkhammar@angstrom.uu.se, phone:
%+46704464271}
%\address{b KTH Royal Institute of Technology \& Stockholm University,
%Roslagstullsbacken 23, SE-106 91, Stockholm, Sweden, lars.mattsson@nordita.se}
%\address{c Department of Mathematics, Uppsala University, SE-
%751 06 Uppsala, Sweden. e-mail: jesper.ryden@math.uu.se.}

\maketitle

\begin{abstract}
Based on recent results from general relativistic statistical mechanics and
black hole information transfer limits a space-time entropy-action equivalence
is proposed as a generalization of the holographic principle.
With this conjecture, the action principle can be replaced by
the second law of thermodynamics, and for the Einstein-Hilbert
action the Einstein field equations are conceptually the result of
thermodynamic equilibrium. For non-equilibrium situations
Jaynes' information-theoretic approach to maximum entropy production
is adopted instead of the second law of thermodynamics. As it turns out, for appropriate
choices of constants quantum gravity is obtained. For the special case of
a free particle the Bekenstein-Verlinde entropy-to-displacement relation
of holographic gravity, and thus the traditional holographic principle, emerges.
Although Jacobson's original thermodynamic equilibrium approach proposed
that gravity might not necessarily be quantized, this particular
non-equilibrium treatment might require it.%the "second entropy" is obtained through maximum entropy production
\end{abstract}

%\begin{keywords}
%Entropy \sep Non-equilibrium thermodynamics \sep Quantum gravity \sep Information theory
%\end{keywords}

%\begin{keyword}
%Probability density distributions \sep Household power consumption \sep  Electric vehicle home-charging \sep Photovoltaic power production
%\end{keyword}

%\end{frontmatter}
%\tableofcontents

\section{Introduction}
Gravity, thermodynamics and quantum mechanics are deeply connected.

%Quantum mechanics and thermodynamics are intimately connected
The partition function for a grand canonical ensemble and the quantum partition function
derived from Feynman's path integral formulation are analogous with the substitution
of time and inverse temperature $i t/\hbar  \leftrightarrow 1/k_b T$ \cite{Lisi}. While such a substitution ---
formally corresponding to a Wick rotation --- might
be useful for computations, it is also a key to connect both theories
via Shannon information theory \cite{Lisi,Acosta,Baez,Munkhammar}.
%which is also the conceptual foundation for the relational interpretation of
%quantum mechanics \cite{Rovelli2}.

%While Wick rotations such as this one may be useful for computations, it is al
%This is useful
%for computations, but it is also possible to make a conceptual
%connection via information theory \cite{Acosta,Baez,Lisi,Munkhammar}.

As regards thermodynamics and gravity,
Jacobson showed that the Einstein field equations could be derived based
on space-time thermodynamic assumptions of Unruh temperature and
Rindler coordinates \cite{Jacobson}. This connection was deepened
with the emergence of the holographic principle from string theory, loop quantum gravity and entropic gravity
theories \cite{Bekenstein2,tHooft,Maldacena,Bousso,Paddy1,Paddy2,Paddy3,Susskind,Verlinde}.
This suggests that entropy --- and thus Shannon information theory  ---
might potentially be an important aspect in the connection between thermodynamics and gravity
as well.
%might potentially be the point of connection between thermodynamics and gravity
%as well. %, albeit with a holographic principle in the current state-of-the art.

However, as regards entropy, a complete covariant theory of thermodynamics and statistical mechanics
in a full general relativistic context is yet to be established \cite{Rovelli}. Initial investigations have shown that for a constant
"thermal time" there seems to be a direct relation between temperature and inverse proper time \cite{Rovelli,Haggard,Cordoba}.
With the assumption that entropy is negative information $I$ there is also a direct connection between
entropy, energy $E$ and time $t$ in upper limits of the bound on the
information $I$ transfer in black hole thermodynamics
\cite{Bekenstein}:
\begin{equation}\label{EqnBekenstein}
I < \gamma E t
\end{equation}
%
% With the definition
%of thermodynamic entropy as $dH = \delta Q/T$ this hints that entropy might be proportional
%to energy and time.
for the constant $\gamma = \pi/\hbar \ln 2$. Although this is in line with the holographic principle,
it is perhaps worthwhile to consider an alternative holographic principle based on these
results. A principle which connects entropy dynamically to energy and time, such that
an equilibrium configuration is a special case, like in \cite{Eling}. Non-equilibrium is after
all believed to be the most common state of a thermodynamics system \cite{Eling,Martyushev}.

\section{Entropy-action equivalence conjecture}
If there is a potential proportionality between inverse temperature and time
the thermodynamic relation  $dH = \delta Q/T$ proposes
a direct proportionality between entropy $H$, energy $E$ and time $t$:
\begin{equation}
H \propto E \cdot t,
\end{equation}
which is up to a constant the the upper information transfer limit for black holes
\eqref{EqnBekenstein} derived by Bekenstein \cite{Bekenstein}. This general entropy is
dependent on coordinate variables of space and time. For most purposes the
second law of thermodynamics is enforced to constrain the dynamics. In effect
this means a maximization of the entropy. Keep in mind that the form of energy
$E$ is not constrained here, and that basically it is set up to match units of the
dynamical components of \eqref{EqnBekenstein} and that of thermal time.

Thus consider, for the time being, the possibility that $E$ is the Lagrangian $L$ of a
system (in classical terms) and that the Lagrangian is integrated over time $t$ to bring the entropy $H$:
\begin{equation}\label{EqnEntropyAction1}
H \propto \int L(q,\dot{q},t) \, dt,
\end{equation}
with a generalized coordinate $q$. This is dramatically different from Bekenstein's information transfer bound,
but the units of the dynamical components remain. And for the special case of $L = E$
and that $E$ is independent of time $t$ brings $H \propto E\cdot t$ which is the upper limit of
Bekenstein's information transfer bound. Also, if one applies the second law of thermodynamics
to \eqref{EqnEntropyAction1}, with respect to the generalized coordinates $q$ and $\dot{q}$, it
would be equivalent to the variational principle. This is worthwhile investigating deeper, since if applicable
it could potentially dispense with the use of certain action principles in physics.

%If applicable it could imply one less principle
%to consider in physics: the action principle.

The most general classical description of physics is given by general relativity, whose dynamics is traditionally
derived from the Einstein-Hilbert action. As the main conjecture of this paper, let's assume, for the time being, that
the space-time thermodynamic entropy $H$ for a space-time region $\mathcal{D}$ can be identified with
the Einstein-Hilbert action $S_{EH}$ with a proportionality constant $\kappa$:

%With this in mind, let the space-time thermodynamic entropy $H$ for a space-time domain $\mathcal{D}$
%be identified with the Einstein-Hilbert action $S_{EH}$ with a proportionality constant $\kappa$:

%classical Einstein-Hilbert action $S_{EH}$ can be identifie
%proportional to the space-time thermodynamic entropy $H$ by a proportionality constant $\kappa$:
%I will here assume that the same holds for the
%classical Einstein-Hilbert action $S_{EH}$:
%
\begin{equation}\label{EqnEH}
H(g_{\mu \nu},\mathcal{L}_M,\mathcal{D}) \equiv \kappa S_{EH}(g_{\mu \nu},\mathcal{L}_M,\mathcal{D})  = \kappa \int_\mathcal{D} \Big(\frac{c^4}{16\pi G} R +\mathcal{L}_M \Big) \sqrt{-g} \, d^4x.
\end{equation}
where $R = R(g_{\mu \nu})$ is the Ricci-scalar based on the metric tensor $g_{\mu \nu}$ and
$\mathcal{L}_M$ is the matter-Lagrangian. The entropy is then a function of the metric $g_{\mu \nu}$
(as a set of generalized coordinates), the classical matter-Lagrangian $\mathcal{L}_M$ and space-time region $\mathcal{D}$.
This means that the entropy is dependent on both space-time configuration and matter content within a
domain of space-time. Entropy, as defined here, is still in practice in proportionality with the dynamical variables of energy and time.
%Then for each matter source $\mathcal{L}_M$ this system
%is defined by the setup of the metric tensor $g_{\mu \nu}$. Also, given $\mathcal{L}_M$ then each configuration
%of $g_{\mu \nu}$ has a defined entropy $H = H(g_{\mu \nu})$.

If entropy $H$ is varied with respect to the generalized coordinates, that is the metric $g_{\mu \nu}$, then
a "thermodynamic equilibrium state" is conceptually obtained. Formally the Einstein field equations
are given, since that variation corresponds to the traditional variation of the Einstein-Hilbert action.
These derivations are formally equivalent, but there is a conceptual difference for the variation:
\emph{maximization of entropy} instead of \emph{traditional variation of the action}, which will be of
importance as the paper progresses.

%which is formally here equivalent to the Einstein field equations.
%This process is formally equivalent to the traditional
%variation of the Einstein-Hilbert action in order to obtain the Einste

%not repeated here. Instead the difference lies in the proposition behind the variation: maximization of entropy %instead of traditional variation of the action.

%In nature, most thermodynamic systems are not
There is no formal novelty in varying the Einstein-Hilbert action
with respect to the metric, but there is a conceptual gain via the
connection to entropy and thermodynamics.
In thermodynamics, equilibrium is not always achieved.
In fact a state of maximum entropy is rarely achieved, and most systems
are likely in a non-equilibrium state \cite{Eling,Martyushev}.
Theoretically, for thermodynamic applications, in non-equilibrium situations
a \emph{maximum entropy production approach} is adopted instead of
the second law of thermodynamics which postulates a \emph{maximum entropy approach}
\cite{Eling,Martyushev}.

%Instead
%effects of non-equilibrium thermodynamics arise. Typically a maximum entropy production
%approach is adopted to obtain the dynamics of the non-equilibrium thermodynamic state.

%From entropic gravity perspective
%it is reasonable that

%The traditional process to obtain space-time dynamics
%of the Einstein-Hilbert action is to perform variation of the action based on the variational
%principle. By assuming a proportionality between entropy and action then this process
%constitutes a variation of entropy, which is instead a purely information-theoretic derivation.
%Note here that $H=H(g_{\mu \nu})$ and that equilibrium corresponds to extremized $H$.
%The Jacobson theory was extended by
%by Eling, Guedens and Jacobson to non-equilibrium space-time thermodynamics in \cite{Eling}.

In Jacobson's original thermodynamic approach to gravity the Einstein field equations were
derived on the assumption of thermodynamic equilibrium  \cite{Jacobson}. This ansatz
was extended to non-equilibrium space-time thermodynamics by Eling, Guedens and Jacobson in \cite{Eling}.
This study assumed that the entropy was a function of the Ricci scalar:
\begin{equation}\label{EqnEling}
H \propto \sigma + f(R).
\end{equation}
As a means to estimate the rate of entropy change semi-classical components such as the Unruh temperature $T$
and "internally developed entropy" $dH_i$ were used in the entropy balance equation
$dH = \delta Q/T + d_i H$. Generally, see \cite{Martyushev} for an overview of entropy production methodologies.
Albeit a semi-classical approach the study in \cite{Eling} revealed interesting connections between shear
viscosity of the horizon in the Einstein field equations and entropy. While the entropy-action equivalence
\eqref{EqnEH} is a function of space and time, it is also --- in similarity with \eqref{EqnEling} a function of the Ricci scalar.
Eling, Guedens and Jacobson's approach regarded a general setup, while the approach here is more specified
and by onset not based on semi-classical concepts. In their study gravity was based on semi-classical
concepts, but remained, however, classic.
%is also a function of the Ricci scalar, albeit with space
%and time as well, and considerably

%The similarity with this approach is that the entropy-action equivalence \eqref{EqnEH} is similar

%This study suggested that the
%entropy-area relation of of was universal, even for non-equilibrium.

%In this study gravity remained, however,
%classic.

An alternative approach for describing non-equilibrium thermodynamics is to utilize
Janyes' information theoretic setup \cite{Jaynes1,Jaynes2,Jaynes3,Jaynes4}. As a macroscopic approach
--- similar to that of thermodynamics itself --- that method devises a partition function based on all possible
states of entropy and uses a maximum entropy production principle to constrain probabilities for
different states. If applied to Eling et al.'s study it would imply setting up a partition function
for the unknown function of the Ricci scalar. Formally in this approach, we have the following.

Assume the proposed space-time entropy of the
classical Einstein-Hilbert action \eqref{EqnEH}, but
resist the hesitation to apply variation to obtain thermodynamic equilibrium.
Instead, setup all possible states of configuration for
$H$ in terms of $g_{\mu \nu}$. Based on these states a new form of entropy
$H_2$ can be identified. In traditional non-equilibrium thermodynamics this is labeled "the second entropy"
\cite{Martyushev,Attard}, which is the pure
information-theoretic entropy of all states of entropy $H[g_{\mu\nu}]$:
\begin{equation}
H_2 = \sum_{All\, g_{\mu\nu}} p[g_{\mu\nu}] \log p[g_{\mu\nu}] = -\int Dg_{\mu\nu} p[g_{\mu\nu}]\log p[g_{\mu\nu}]
\end{equation}
where $p[g_{\mu\nu}]$ is an unknown probability function defined for each configuration of $g_{\mu\nu}$.
This produces expected values, for example for the ("first") entropy $H$:
\begin{equation}\label{ExpectedEntropy}
\langle H \rangle = \sum_{All \, g_{\mu\nu}} p[g_{\mu\nu}] H[g_{\mu\nu}] = \int Dg_{\mu\nu} p[g_{\mu\nu}]H[g_{\mu\nu}],
\end{equation}
which is proportional to the expected action via \eqref{EqnEH}.
The probability function $p[g_{\mu\nu}]$ is found by using Lagrange multipliers, as was
done for non-equilibrium thermodynamics in \cite{Jaynes1,Jaynes2,Jaynes3,Jaynes4} and
analogously for information-theoretic interpretations of quantum mechanics in
\cite{Lisi,Baez,Munkhammar,Lee}.
%In \cite{Lisi} Lisi performed the following derivation
%which is worth repeating here.
By employing complex Lagrange multipliers,
$\lambda$ and $\alpha$, the second entropy $H_2$ %\eqref{Entropy}
is maximized in terms of "first" entropy $H$ by:
\begin{equation*}
H_2 = - \int Dg_{\mu\nu} p[g_{\mu\nu}] \log p[g_{\mu\nu}] + \lambda \Big(1-\int Dg_{\mu\nu}p[g_{\mu\nu}]\Big) +
\end{equation*}
\begin{equation}
+ \alpha \Big(\langle H[g_{\mu\nu}] \rangle - \int Dg_{\mu\nu} p[g_{\mu\nu}]H[g_{\mu\nu}]\Big),
\end{equation}
which simplified becomes:
\begin{equation}
H_2 = \lambda + \alpha \langle H\rangle - \int Dg_{\mu \nu} (p[g_{\mu\nu}]\log p[g_{\mu\nu}] +
\lambda p[g_{\mu\nu}] + \alpha p[g_{\mu\nu}] H[g_{\mu\nu}]).
\end{equation}
If we perform variation on the probability distribution we get:
\begin{equation}
\delta H_2 = - \int Dg_{\mu\nu} (\delta p[g_{\mu\nu}])(\log p[g_{\mu\nu}] + 1 + \lambda +
\alpha H[g_{\mu\nu}])
\end{equation}
which is extremized when $\delta H_2 = 0$, which corresponds to the
probability distribution:
\begin{equation}\label{Probability}
p[g_{\mu \nu}] = e^{-1-\lambda} e^{-\alpha H[g_{\mu\nu}]} = \frac1Z e^{-\alpha H[g_{\mu\nu}]}.
\end{equation}
%
%which is compatible with the knowledge constraints \cite{Lisi}.
By varying the Lagrange multipliers one enforces the two constraints,
giving $\lambda$ and its connection to $\alpha$. Especially one gets:
$e^{-1-\lambda} = \frac1Z$ where $Z$ is a form of partition function:
\begin{equation}\label{Partition}
Z = \sum_{g_{\mu\nu}} e^{-\alpha H[g_{\mu\nu}]} = \int Dg_{\mu\nu} e^{-\alpha H[g_{\mu\nu}]}.
\end{equation}
This yields the expected entropy:
\begin{equation}%\label{ExpectedEntropy}
\langle H \rangle = \sum_{g_{\mu\nu}} p[g_{\mu\nu}] H[g_{\mu\nu}] = \sum_{g_{\mu\nu}} \frac1Z e^{-\alpha H[g_{\mu\nu}]} H[g_{\mu\nu}]
= \int Dg_{\mu\nu} \frac1Z e^{-\alpha H[g_{\mu\nu}]} H[g_{\mu\nu}],%\int Dq p[q]H[q].
\end{equation}
which by \eqref{EqnEH} is proportional to the expected action with proportionality constant $\kappa$:
\begin{equation*}%\label{ExpectedEntropy}
\langle S \rangle = \kappa \sum_{g_{\mu\nu}} p[g_{\mu\nu}] S_{EH}[g_{\mu\nu}] = \kappa \sum_{g_{\mu\nu}} \frac1Z e^{-\alpha \kappa S_{EH}[g_{\mu\nu}]}  S_{EH}[g_{\mu\nu}] =
\end{equation*}
\begin{equation}
= \kappa \int Dg_{\mu\nu} \frac1Z e^{-\alpha \kappa S_{EH}[g_{\mu\nu}]} S_{EH}[g_{\mu\nu}].%\int Dq p[q]H[q].
\end{equation}
The second entropy --- also called the "entropy of entropy" arises in Jaynes' non-equilibrium formulation
here and is defined as:
%It is also called "second entropy", and is here defined as:
%The "entropy of entropy", or "second entropy", is then:
%
\begin{equation}%\label{ExpectedEntropy}
H_2 = - \sum_{g_{\mu\nu}} p[g_{\mu\nu}] \log(p[g_{\mu\nu}]) = - \int Dg_{\mu\nu} p[g_{\mu\nu}]\log([g_{\mu\nu}]),
\end{equation}
which by the criterion $\sum_{All \, g_{\mu\nu}} p[g_{\mu\nu}] = 1$ and algebraic manipulations become
(see derivation in \cite{Baez,Munkhammar}):
\begin{equation}
H_2 = - \alpha \langle H \rangle + \log Z.
\end{equation}
%
%This suggests that for situations which are closer to equilibrium,
%where all probabilities approach zero except one, the second entropy is only the factor $-\alpha$
%compared with the "first" entropy:
%
%\begin{equation}
%H_2 ~ - \alpha H.
%\end{equation}
%

This characterizes the complete description of possible states of entropy and second
entropy in this setup, with undefined constants $\kappa$ connecting action and entropy,
and $\alpha$ related to entropy production maximization. Sections \ref{SectionQuantum}
and \ref{SectionHolographic} will outline the pursuit of these constants.
%The pursuit for these constants
%will

In short, this approach represents the complete non-equilibrium dynamics of general relativity based on
the action-entropy equivalence conjecture and Jayne's second entropy approach for maximum
entropy production.
%and $\alpha$.

\section{$\alpha \kappa$ and quantum gravity}\label{SectionQuantum}
In this setup two constants are undefined so far: $\kappa$ which connects entropy $H$
to the action $S$, and $\alpha$ which connects entropy $H$ to second entropy $H_2$.
By replacing $H$ with $\kappa S$ in the partition function, according to \eqref{EqnEH},
and assuming that $\alpha \kappa \equiv  i/\hbar$ then one gets:
%*one gets:
%
%\begin{equation}\label{Partition}
%Z = \sum_{q} e^{-\alpha \kappa S[q]} = \int Dq e^{-\alpha \kappa S[q]}.
%\end{equation}
%
%This is strikingly similar to the quantum mechanical partition function:
%
\begin{equation}\label{Partition}
Z = \sum_{g_{\mu\nu}} e^{\frac{i}{\hbar} S[g_{\mu\nu}]} = \int Dg_{\mu\nu} e^{\frac{i}{\hbar} S[g_{\mu\nu}]},
\end{equation}
which is the quantum partition function for quantum gravity \cite{Gibbons}. This also renders
the second entropy equivalent to an information-theoretic entropy in quantum mechanics, also called
"quantropy", as the term was coined by Baez and Pollard \cite{Baez,Munkhammar}.

Since this connects to information theory, and previous studies on information theoretic approaches
to quantum mechanics, an explicit probability for each configuration of $g_{\mu \nu}$ is provided
(seen for general configurations in \cite{Lisi,Baez,Munkhammar}):
%In addition to providing a partition function for quantum dynamics
%this information theoretic setup also
%provides a form of probability for each configuration $g_{\mu\nu}$ (also seen in
%\cite{Lisi,Munkhammar,Baez}):
%
\begin{equation}\label{EqnProbability2}
p[g_{\mu \nu}] =  \frac1Z e^{-\frac{i}{\hbar} S[g_{\mu\nu}]}.
\end{equation}
Because $\alpha \kappa =  i/\hbar$ is complex the probability for each state is
complex, which is problematic from probability theory perspective. However, that can be mended
with the following setup, which was devised by Lisi for general quantum systems \cite{Lisi}. The probability
for the system to be on a specific path in a set of possible paths in
configuration space is:
\begin{equation}
p(set) = \sum_{paths} \delta_{path}^{set} p[path] = \int Dg_{\mu \nu}
\delta(set-g_{\mu \nu})p[g_{\mu \nu}].
\end{equation}
Typically the system reverses sign under inversion of parameter integration
limits \cite{Lisi}:
%Here Lisi assumed that the action typically reverses sign under
%inversion of the parameters of integration limits:
%
\begin{equation}
S^{t'} = \int^{t'} dt \, L(g_{\mu \nu}) = - \int_{t'} dt \, L(g_{\mu \nu}) =
- S_{t'}.
\end{equation}
This implies that the probability for the system to pass through
configuration $g_{\mu \nu}'$ at parameter value $t'$ is:
\begin{equation*}
p(g_{\mu \nu}',t') = \int Dg_{\mu \nu} \delta(g_{\mu \nu}(t') -g_{\mu \nu})p[g_{\mu \nu}] =
\end{equation*}
\begin{equation*}
= \Bigg(\int^{g_{\mu \nu}(t')=g_{\mu \nu}'}
Dg_{\mu \nu}p^{t'}[g_{\mu \nu}] \Bigg) \Bigg(\int_{g_{\mu \nu}(t')=g_{\mu \nu}'}
Dg_{\mu \nu}p_{t'}[g_{\mu \nu}] \Bigg) =
\end{equation*}
\begin{equation}\label{PsiProbability}
= \psi(g_{\mu \nu}',t')\psi^\dagger(g_{\mu \nu}',t'),
\end{equation}
in which we can identify the quantum wave function:
\begin{equation*}
\psi = \int^{g_{\mu \nu}(t')=g_{\mu \nu}'} Dg_{\mu \nu} p^{t'} [g_{\mu \nu}] = \frac{1}{\sqrt{Z}}
\int^{g_{\mu \nu}(t')=g_{\mu \nu}'} Dg_{\mu \nu} e^{-\alpha S^{t'}} =
\end{equation*}
\begin{equation}
= \frac{1}{\sqrt{Z}}
\int^{g_{\mu \nu}(t')=g_{\mu \nu}'} Dg_{\mu \nu} e^{i\frac{S^{t'}}{\hbar}}.
\end{equation}
This gives the probability amplitude for the probability of a system to
pass through metric $g_{\mu \nu}'$ at time $t'$. See the similarity with
quantum mechanics in \cite{Lisi}.

%The quantum wave function $\psi(g_{\mu \nu}',t)$ defined here is valid for
%paths $t<t'$ meeting at $g_{\mu \nu}'$ while its complex conjugate
%$\psi^\dagger(g_{\mu \nu}',t')$ is the amplitude of paths with $t>t'$
%leaving from $g_{\mu \nu}'$. Multiplied together they bring the probability
%amplitudes that gives the probability of the system passing
%through $g_{\mu \nu}'(t')$, as is seen in \eqref{PsiProbability}.

It should be noted that that the quantum wave function in quantum
mechanics is subordinate to the partition function since it only works
when $t'$ is a physical parameter of the system and that the system
is $t'$ symmetric, which provides a real partition function $Z$.
%However,
%just as Lisi points out \cite{Lisi}, this quantum wave function in
%quantum mechanics is subordinate to the partition function
%formulation since it only works when $t'$ is a physical parameter
%and the system is $t'$ symmetric, providing a real partition
%function $Z$.

Thus, with appropriate constants it is perhaps unexpectedly possible to obtain quantum
gravity from this non-equilibrium ansatz. However, only the product
of constants $\alpha \kappa$ have been determined so far. In order to
obtain the equivalence proportionality constant $\kappa$ a connection
to the holographic principle is made in the following section.

%In fact it, surprisingly, also
%connects to the holographic principle, as is shown in the following section.

%A real version of this probability can be obtained by applying
%Wick-rotation of time in the action.

\section{$\alpha$ and holographic gravity}\label{SectionHolographic}
%In the
The entropy-action equivalence conjecture with the Einstein-Hilbert action here proposes
that the action principle is equivalent to the second law of thermodynamics. That is, classical
dynamics of a system is obtained for thermodynamic equilibrium situations and quantum
dynamics arise in non-equilibrium situations. However all of this requires that the entropy-action
equivalence conjecture in fact is valid and indeed supersedes the holographic principle.

A first step on that path is to show equivalence under certain conditions.
According to the entropy-action equivalence conjecture \eqref{EqnEH} the entropy of a stationary particle
with mass $m$ is:% sufficiently far away from it so that the Ricci-scalar $R\sim 0$ becomes:
% is highlighted
%as a generalization of the holographic principle
%In this paper space-time entropy is proposed to be proportional to the Einstein-Hilbert action.
%With the assumption of an entropy-action equivalence \eqref{EqnEH} the entropy of a stationary particle
%with mass $m$ is defined by:
%
\begin{equation}
H = \kappa \int  mc^2 dt = \kappa mc^2 t.
\end{equation}
%\begin{equation}
%S \sim \kappa \int mc^2 dt = \kappa mc^2 t.
%\end{equation}
%
If we assume that time can be expressed as $\Delta x/c$ for a "time-like" distance
$\Delta x$ then we get:
\begin{equation}
\Delta S = \kappa mc \Delta x,
\end{equation}
and if we set $\kappa = - 2\pi k_B/\hbar$, where $k_B$ is Boltzmann's constant,
then we get the Bekenstein-Verlinde expression from holographic gravity \cite{Verlinde}:
\begin{equation}\label{EqnBekensteinVerlinde}
\Delta S = -2\pi k_B \frac{mc}{\hbar} \Delta x.
\end{equation}
This expression was originally derived by Bekenstein (up to a constant) \cite{Bekenstein} for particles
falling into black holes, and it was more recently slightly altered by Verlinde
to develop his approach to entropic gravity \cite{Verlinde}.
If this special case of the entropy can be assumed
equivalent to the Bekenstein-Verlinde expression it is possible to
identify $\alpha$ as well since $\alpha \kappa = \frac{i}{\hbar}$:
%Incidentally since $\alpha \kappa = \frac{i}{\hbar}$ this also identifies $\alpha$:
%
\begin{equation}
\alpha = \frac{i}{\hbar \kappa} = -\frac{i}{2\pi k_B}.
\end{equation}
This defines both quantum dynamics in \eqref{Partition} and the entropy-action proportionality
for the Einstein-Hilbert action $S_{EH}$ \eqref{EqnEH}:
\begin{equation}\label{EqnEH2}
H  \equiv -\frac{2\pi k_b}{\hbar}  S_{EH},
\end{equation}
which completes the investigation on the generalization of the holographic principle
by means of an entropy-action equivalence conjecture in this paper.

\section{Discussion and conclusions}
Based on recent developments in general relativistic statistical mechanics and black hole
thermodynamics an entropy-action equivalence is conjectured. With it, Einsteinian
gravity is obtained for equilibrium situations --- analogous to Jacobson's results ---
and a non-equilibrium theory of gravity is also developed based on Jaynes'
information-theoretic approach to maximum entropy production.

This approach to maximum entropy production, in thermodynamics and other fields,
is one among several different approaches to characterize non-equilibrium
dynamics, and the non-equilibrium concept is not without controversy \cite{Martyushev}.
But the fact that for a specific choice of the coupling constant and a Lagrange
multiplier the result is quantum gravity is perhaps interesting enough to consider
this possibility. And if true, it also presents a form of information-theoretic quantization
principle: maximum entropy production. Another interesting feature is also that the
correspondence to classical physics is obtained by enforcing the second law of
thermodynamics, which dispenses with the use of a classical action principle.

%Based on the conjecture of an entropy-action equivalence % (instead of the holographic principle)
%an information-theoretic approach to quantum gravity is developed. Jaynes' information
%theoretic approach to maximum entropy production for non-equilibrium thermodynamics
%is used to quantize the dynamics.
%This shows that in this setup --- given appropriate constants --- the thermodynamic equilibrium
%state is satisfied by the Einstein field equations and the non-equilibrium state is represented by
%quantum gravity.

%This also incidentally proposes a form of information-theoretic quantization principle: maximum entropy production.
One advantage of this approach is the derivation of an explicit probability for each
potential metric configuration, which could perhaps be useful for certain calculations \cite{Lisi}.
The entropy-action equivalence is conjectured as a generalization of the holographic
principle and shown to correspond to the Bekenstein-Verlinde entropy-to-displacement relation
of holographic gravity for the special case of a stationary mass. That being said, this
conjecture likely violates the holographic principle, even under classical considerations, for
non-stationary situations.
%Consider the following example.

%The holographic principle postulates that information is stored on two dimensional surfaces. The
%entropy-action equivalence conjecture on the other hand proposes that it is instead inherent to 
%space-time regions, even if it can be 

%It should be noted that while the holographic principle regards information as
%stored on two dimensional surfaces this entropy-action equivalence conjecture
%instead proposes that it is instead inherent to space-time regions.

The entropy of any object, even a black hole, is universal for all observers in this approach (since the classical
Lagrangian is universal) and based on the energy and matter content of space-time.
This information-theoretic quantization of gravity suggests that space-time is encoded with
the fundamental stochastic nature of quantum mechanics. Even though entropy is invariant
for all observers, the second entropy: quantropy, is not. Since observers may or may not gain
information regarding the state of any object, quantropy is by necessity observer
dependent \cite{Lisi,Munkhammar}.
%This quantity is by necessity observer dependent.

%the stochastic variable here,
%Instead, it is space-time that is fundamentally unknown, and in a non-equilibrium situations
%the result is a second entropy: quantropy. Also, the quantropy is --- unlike the entropy ---
%observer dependent.

%The principle of general covariance proposes that the form of physical laws
%is invariant under arbitrary differentiable coordinate transforms.

In a foundational sense, since the theory is based on accessible information and
observer dependence, it seems to demand some form of \emph{principle of information covariance}:
the laws of physics can only be defined on the basis of the information accessible
to each observer \cite{Munkhammar}. This suggests that perhaps Rovelli's relational
interpretation for quantum mechanics is favorable for this theory \cite{Rovelli2}.

\section*{Acknowledgements}
The author wishes to thank Dr. Garrett Lisi, Dr. Ingemar Bengtsson, Dr. Lars Mattsson and Dr. Jacob Bekenstein for valuable
comments on the theory and the paper. %potential limits of the theory.

%The author wishes to thank Dr. Garrett Lisi for valuable discussions
%on information theory and quantum mechanics.

\bibliographystyle{elsarticle-num}

\begin{thebibliography}{00}

\bibitem{Lisi}
Lisi, G.: Quantum mechanics from a universal action reservoir. arXiv:physics/0605068v1 (2006)

\bibitem{Acosta}
Acosta, D., de Cord\'{o}ba,~P.~F., Isidro,~J.~M., Santander,~J.~L.~G.:
A holographic map of action onto entropy. Journal of Physics:
Conference Series 361, pp. 1-9, (2012)

 \bibitem{Baez}
Baez,~J.~C.,  Pollard,~B.~S.: Quantropy, Entropy 17: 772-789, arXiv:1311.0813 (2013)

\bibitem{Munkhammar}
Munkhammar,~J.~D.: Canonical Relational Quantum Mechanics from Information Theory.
EJTP 8:93-108, arXiv:1101.1417 (2011)

\bibitem{Jacobson}
Jacobson,~T.: Thermodynamics of Spacetime: The Einstein Equation of State.
Phys.Rev.Lett.75, pp. 1260-1263, arXiv:gr-qc/9504004v2 (1995)

\bibitem{Bekenstein2}
Bekenstein,~J.~D.: Black holes and entropy. Phys. Rev. D. 7: pp. 2333-2346 (1973)

\bibitem{tHooft}
't Hooft,~G.: Dimensional reduction in quantum gravity.
THU-93/26, arXiv:gr-qc/9310026v2 (1993)

\bibitem{Maldacena}
Maldacena,~J.: The Large N Limit of Superconformal Field Theories and Supergravity.
Adv.Theor.Math.Phys.2, pp. 231-252, arXiv:hep-th/9711200 (1998)

\bibitem{Bousso}
Bousso,R.: The holographic principle. Rev.Mod.Phys.74, pp. 825-874,
arXiv:hep-th/0203101v2 (2002)

\bibitem{Paddy1}
Padmanabhan,~T.: Equipartition of energy in the horizon degrees of freedom and the
emergence of gravity, Mod.Phys.Lett.A25, pp.1129-1136, arXiv:0912.3165. (2010)

\bibitem{Paddy2}
Padmanabhan,~T.: Gravitational Entropy Of Static Space-Times And Microscopic Density Of
States. Class. Quant. Grav. 21, pp. 4485-4494, arXiv:gr-qc/0308070 (2004)

\bibitem{Paddy3}
Padmanabhan,~T.: Thermodynamical Aspects of Gravity: New insights.
Rep. Prog. Phys. 73, arXiv:0911.5004v2[gr-qc] (2010)

\bibitem{Susskind}
Susskind,~L.: The World as a Hologram. J.Math.Phys.36, pp. 6377-6396,
arXiv:hep-th/9409089v2 (1995)

\bibitem{Verlinde}
Verlinde,~E.: On the Origin of Gravity and the Laws of Newton.
JHEP 1104:029, arXiv:1001.0785v1 [hep-th] (2011)

\bibitem{Rovelli}
Rovelli~C.: General relativistic statistical mechanics.
Phys. Rev. D 87, 084055, arXiv:1209.0065v2 [gr-qc] (2013)

\bibitem{Haggard}
Haggard,~H.~M., Rovell,~C.: Death and resurrection of the
zeroth principle of thermodynamics. Phys. Rev. D 87,
arXiv:1302.0724 [gr-qc] (2013)

\bibitem{Cordoba}
De C\'{o}rdoba,~P.~F., Isidro,~J.~M. , Perea,~M.~H.:
Emergent quantum mechanics as a thermal ensemble.
International Journal of Geometric Methods in Modern
Physics 11. (2014)
%Int. J. Geom. Methods Mod. Phys. 11
\bibitem{Bekenstein}
Bekenstein,~J.~D.: Energy Cost of Information Transfer.
Phys. Rev. Lett. 46, pp. 623-626 (1981)

 \bibitem{Eling}
Eling,~C., Guedens,~R., Jacobson,~T.: Non-equilibrium Thermodynamics of
Spacetime. Phys.Rev.Lett.96:121301, arXiv:gr-qc/0602001v1 (2006)

\bibitem{Martyushev}
Martyushev,~L.~M., Seleznev,~V.~D.: Maximum entropy production principle in
physics, chemistry and biology. Physics Reports 426: pp. 1-45 (2006)

\bibitem{Jaynes1}
Jaynes,~E.~T.: Information theory and statistical mechanics. Physical Review 106, pp.
620-630. (1957)

\bibitem{Jaynes2}
Jaynes,~E.~T.: Information theory and statistical mechanics II. Physical Review 108:
pp. 171-190. (1957)

\bibitem{Jaynes3}
Jaynes,~E.,~T.: Macroscopic prediction, in Complex Systems - Operational Approaches in
Neurobiology. Edited by H. Haken, Springer-Verlag, Berlin, pp. 254-269, ISBN 3-540-15923-1 (1985).

\bibitem{Jaynes4}
Jaynes,~E.~T.: Gibbs vs Boltzmann Entropies, American Journal of Physics 33,
pp. 391-398. (1965)

\bibitem{Attard}
Attard,~P.: The Second Entropy: A Variational Principle for
Time-dependent Systems. Entropy 10, pp. 380-391. (2008)

\bibitem{Lee}
Lee~J.-W.: On the Origin of Entropic Gravity and Inertia.
Found. Phys. 42, pp.1153-1164. (2012)

\bibitem{Gibbons}
Gibbons,~G.,~W., Hawking~S.~W.: Action integrals and partition functions
in quantum gravity. Phys. Rev. D, 15, pp. 2752-2756. (1977)

\bibitem{Rovelli2}
Rovelli,~C.: Relational Quantum Mechanics, Int. J. of Theor. Phys. 35, pp. 1637-1678,
arXiv:quant-ph/9609002 (1996)


%\bibitem{Thorn}
%C. B. Thorn, Reformulating string theory with the 1/N expansion,
%International A.D. Sakharov Conference on Physics. Moscow. pp. 447�454.
%arXiv:hep-th/9405069. ISBN 978-1-56072-073-7.


\end{thebibliography}

\end{document}